\documentclass[a4paper,fleqn,usenatbib]{mnras}

\usepackage[T1]{fontenc}
\usepackage{ae,aecompl}

\usepackage{graphicx}	
\usepackage{amsmath}	
\usepackage{amssymb}	
\usepackage{color}
\usepackage{float}

\definecolor{grey}{rgb}{0.7,0.7,0.7}

\newcommand{\bluetides}{{\sc BlueTides}}

\newcommand{\Msun}{\rm{M_{\odot}}}
\newcommand{\wfirst}{{\em WFIRST}}

\title[Monsters in the Dark]{Monsters in the Dark: Predictions for Luminous Galaxies in the Early Universe from the {\sc BlueTides} Simulation}

\author[Dacen Waters et al.]
{Dacen Waters$^{1}$, 
Stephen M. Wilkins$^{2}$,
Tiziana Di Matteo$^{1}$,
Yu Feng$^{3}$, 
\newauthor Rupert Croft$^{1}$,
Daisuke Nagai,$^{4,5}$
\\
$^1$\,McWilliams Center for Cosmology, Physics Dept., Carnegie Mellon
University, PittsburghPA, 15213, USA \\
$^2$\,Astronomy Centre, Department of Physics and Astronomy, University of Sussex, Brighton, BN1 9QH, UK \\
$^3$\,Berkeley Center for Cosmological Physics, University of California at
Berkeley, Berkeley, CA 94720, USA \\
$^4$\,Department of Physics, Yale University, New Haven, CT 06520, USA \\
$^5$\,Yale Center for Astronomy and Astrophysics, Yale University, New Haven, CT 06511, USA
}

\date{Sumbitted to MNRAS on March 31, 2016}

\pubyear{2016}

\begin{document}

\label{firstpage}
\pagerange{\pageref{firstpage}--\pageref{lastpage}}
\maketitle

\begin{abstract}

Using deep {\em Hubble} and {\em Spitzer} observations Oesch et al. (2016) have identified a bright ($M_{\rm UV}\approx -22$) star forming galaxy candidate at $z \approx 11$. The presence of GN-$z11$ implies a number density $\sim 10^{-6}\,{\rm Mpc^{-3}}$,  roughly an order of magnitude higher than the expected value based on extrapolations from lower redshift.  Using the unprecedented volume and high resolution of the \bluetides\ cosmological hydrodynamical simulation, we study the population of luminous rare objects at $z > 10$. The luminosity function in \bluetides\ implies an enhanced number of massive galaxies, consistent with the observation of GN-$z11$.  We find about 30 galaxies at $M_{\rm UV}\approx -22$ at $z = 11$ in the \bluetides\ volume, including a few objects about 1.5 magnitudes brighter. The probability of observing GN-$z11$ in the volume probed by Oesch et al. (2016) is $\sim 13$ per cent. The predicted properties of the rare bright  galaxies at $z = 11$ in \bluetides\ closely match those inferred from the observations of GN-$z11$.  \bluetides\ predicts a negligible contribution from faint AGN in the observed SED. The enormous increase in volume surveyed by \wfirst\ will provide observations of $\sim1000$ galaxies with $M_{\rm UV} < -22$ beyond $z = 11$ out to $z = 13.5$.

\end{abstract}

\section{Introduction}

Galaxies at high-redshift can be identified by taking advantage of the
strong spectral break caused by neutral hydrogen in the intergalactic
medium. By combining observations from {\em Hubble} and {\rm Spitzer}
it is possible to extend this technique to $z\sim 10$ and beyond, with
the first, albeit small, samples now identified (e.g. Oesch et al.\
2012, 2013, 2014, 2015, 2016; Zheng et al.\ 2012; Ellis et al.\ 2013;
Bouwens et al.\ 2015, 2016; Zitrin et al.\ 2014; Ishigaki et al.\
2015).

Most recently, Oesch et al. (2016, hereafter O16) identified a single
bright ($M_{\rm UV}\approx -22$) source (GN-$z11$) located at $z\approx 11$
based on both a photometric and spectroscopic continuum
break. Potential low-redshift contaminants (e.g. passively evolving
galaxies or extreme emission line sources) are ruled out at high
significance. The presence of such a bright object in the early
Universe identified within a relatively small volume is surprising
with an inferred density around an order of magnitude larger than
extrapolations from lower-redshift.

Theoretical predictions for these early times are lacking,
particularly those with the dynamic range to cover both the formation
of individual objects and make large scale statistical studies of
them. A trend in galaxy formation studies has been to carry out
``zoomed'' simulations (e.g., Governato et al. 2007; Agertz \&
Kravtsov 2015, Hopkins et al. 2016), where the full physics algorithms
are only brought to bear in very small subvolumes of larger
simulations. This approach is well suited to modeling observations in
small fields of view, but not when quantitative information on
luminosity functions, clustering statistics or abundance of rare
objects or phenomena is required. Even the high resolution volumes of
the recent cosmological hydrodynamical simulation {\it Illustris}
(Vogelsberg et al. 2014), {\it EAGLE} (Schaye et al. 2015) or
{\it MassiveBlackII} (Khandai et al. 2015) are not sufficient to probe the
high mass/bright end of mass/luminosity function at high redshift.

In this analysis we investigate whether objects like GN-$z11$ are
consistent with those found in   \bluetides\
(Feng et al. 2015, 2016),
the only cosmological hydrodynamic 
simulation so far with sufficient volume, mass and spatial
resolution.
Our analysis is organized as follows: in
Section \ref{sec:BT} we briefly introduce the \bluetides\
simulation. In Section \ref{sec:demo} we investigate the predicted
volume density of bright sources at very high redshift and ascertain
whether the existence of GN-$z11$ is in tension with the model. In 
Section \ref{sec:properties} we investigate the properties of
bright galaxies at very-high redshift. Finally, in Section \ref{sec:c}
we present our conclusions.

\section{The BlueTides Simulation}\label{sec:BT}

The \bluetides\ simulation (see Feng et al. 2015, 2016 for a full
description) was carried out using the Smoothed Particle Hydrodynamics
code {\sc MP-Gadget} with $2\,\times\, 7040^{3}$ particles using the
Blue Waters system at the National Center for Supercomputing
Applications. The simulation evolved a $(400/h)^{3}\,{\rm cMpc^3}$
cube to $z=8$ by which time it contained approximately 200 million
objects (of which 160,000 have stellar masses greater than
$10^{8}\,{\rm M_{\odot}}$). At $z=12$ the number of objects identified
falls to around 20 million, and only around 700 have stellar masses
greater than $10^{8}\,{\rm M_{\odot}}$. The galaxy stellar mass and
rest-frame UV luminosity functions predicted by the simulation (see
{Feng et al. 2016}; {Wilkins et al. 2016}) match observational
  constraints available at $z\approx 8$ (e.g. Bouwens, Oesch, Labbe 2015a; Song, Finkelstein, Ashby 2015).

Galaxy spectral energy distributions (SED) were calculated by coupling the
simulation with the {\sc Pegase} (Fioc \& Rocca-Volmerange 1997)
stellar population synthesis (SPS) model assuming a Chabrier (2003)
initial mass function (IMF) combined with reprocessing by both 
gas and dust (see Wilkins et al., 2016 for details).


\begin{figure}
\centering
\includegraphics[width=20pc]{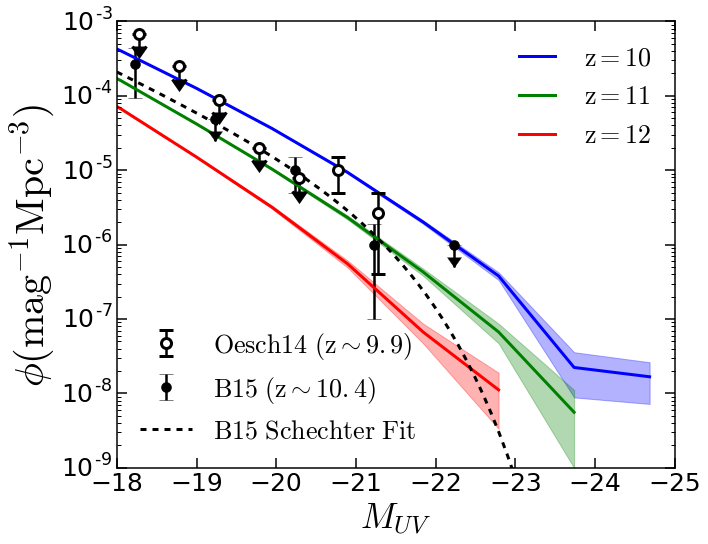}
\caption{Intrinsic rest-frame UV luminosity function (LF) at $z\in\{10,11,12\}$ predicted by the \bluetides\ simulation alongside observational constraints at $z\approx 10.4$ from Bouwens et al. (2015b) and at $z\sim 9.9$ from 
Oech et al. (2014). The shaded regions on the curves
represent the 1$\sigma$ uncertainty on the LF computed from
the entire \bluetides\ volume. }
\label{fig:UVLF}
\end{figure}

\begin{figure}
\centering
\includegraphics[width=21pc]{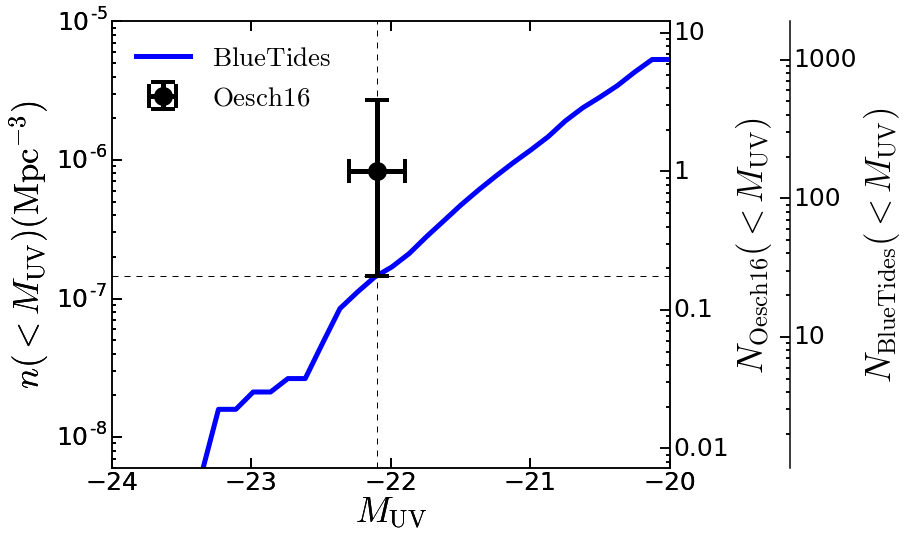}
\caption{The cumulative volume density of galaxies at $z=11$ as
a function of absolute magnitude. 
The observation of GN-$z11$ is shown as a
point, with a horizontal error
bar denoting the uncertainty in the magnitude.
The vertical error is estimated using a simple Poisson error.
The right hand axes show the number of galaxies in both the 
\bluetides\ volume and the volume probed by the O16 analysis.}
\label{fig:volume_density}
\end{figure}

\section{Luminosity Function and Volume Density}\label{sec:demo}
We begin by investigating the predicted luminosity function of
galaxies at $z>10$. In Figure \ref{fig:UVLF} we present the intrinsic
rest-frame UV LF of galaxies at $z=10-12$. The
intrinsic rest-frame UV luminosities, and thus the model expectations, are
sensitive to various assumptions including the choice of IMF, SPS
model, and Lyman continuum escape fraction. The impact of these
assumptions on the SED of galaxies at
high-redshift are explored in detail in Wilkins et al. (2016). Figure \ref{fig:UVLF} shows that observational constraints at $z\approx 10.4$ from Bouwens et al. (2015a) and Oesch et al. (2015) are consistent with these predictions, largely lying between the $z=10$ and $z=11$ lines.

We also show (as a dashed line) the Bouwens et al. (2015a) Schechter fit to
the observed LF at $z=10.4$. For galaxies brighter than $M_{\rm UV} = -22$,
the fit is currently unconstrained by observational data.
\bluetides\ predicts a significantly
larger number of objects than the extrapolation of the
B15 fit into the bright end. 
By interpolating between the $z=11$ and $z=10$ curves
we find that at $z=10.4$ This enhancement is a factor of 5 for 
$M_{\rm UV}=-22$ and $150$ for $M_{\rm UV}=-23$.

To compare directly with the observational results by O16, 
Figure \ref{fig:volume_density} shows the
cumulative volume density of sources at $z=11$. The vertical error bars on the O16 measurement are the 68 per cent exact Poisson confidence intervals. Note that the uncertainties on the number density are simple Poisson errors and do not include additional sources of uncertainties, such as cosmic variance, so that they are likely to be under estimated.
Within the volume probed by O16 ($1.2\times 10^{6}\,{\rm Mpc^3}$) \bluetides\ predicts
approximately $0.17$ objects with intrinsic UV luminosities greater
than $M_{\rm UV}=-22.1$. 

The probability of observing one or more object given
this expectation is 16 per cent assuming a Poisson distribution. 
Since we have the full simulation data, we can find this probability without assuming Poisson statistics. The O16 volume fits into the \bluetides\ volume 156 times, so we subdivide the volume into boxes where one side has a length equal to the comoving distance between $z=10.5 -11.5$. We find in these subvolumes, 17 contain one galaxy with  $M_{\rm UV}<-22.1$. Three of the subvolumes contain two galaxies and one contains four galaxies. This translates to a 13 per cent chance of observing one or more $M_{\rm UV}<-22.1$ galaxies in the O16 observation volume. These very bright objects are very highly biased, with a linear bias of $b\sim 20$ (Waters et al. 2016).

These predictions are based on the intrinsic luminosities and do not include dust attenuation.
Significant dust attenuation in very bright galaxies at $z=11$ would reduce the number density
of (UV bright) sources reducing the otherwise good agreement with the O16
result.


\begin{figure}
\centering
\includegraphics[width=20pc]{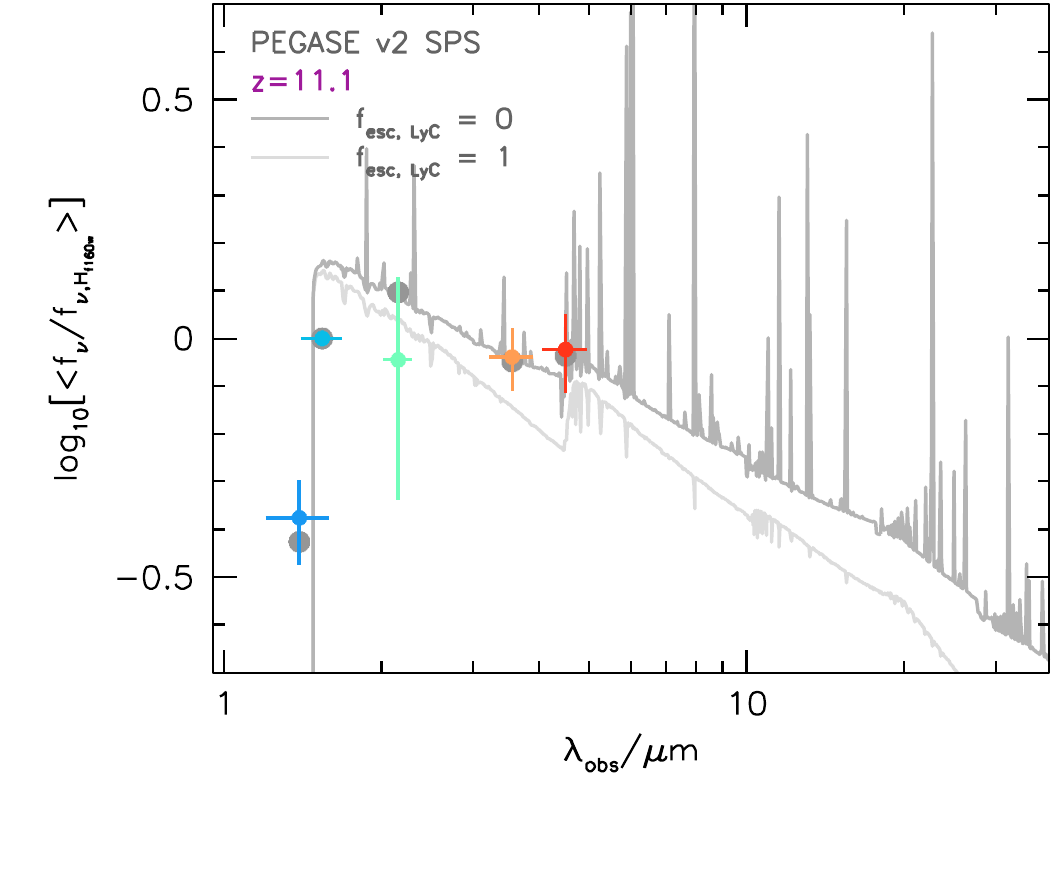}
\caption{The average predicted SED of bright
  ($M_{\rm UV}\approx -22$) galaxies at $z=11.1$
in \bluetides\ compared with the
  observed fluxes of GN-$z11$. The two SEDs
  shown denote both the pure stellar case ($f_{\rm esc, LyC}=1$) and the
  case in which the Lyman continuum escape fraction is effectively zero. 
  In the latter case the Lyman-$\alpha$ line has also been damped.}
\label{fig:SED}
\end{figure}

\begin{figure}
\centering
\includegraphics[width=21pc]{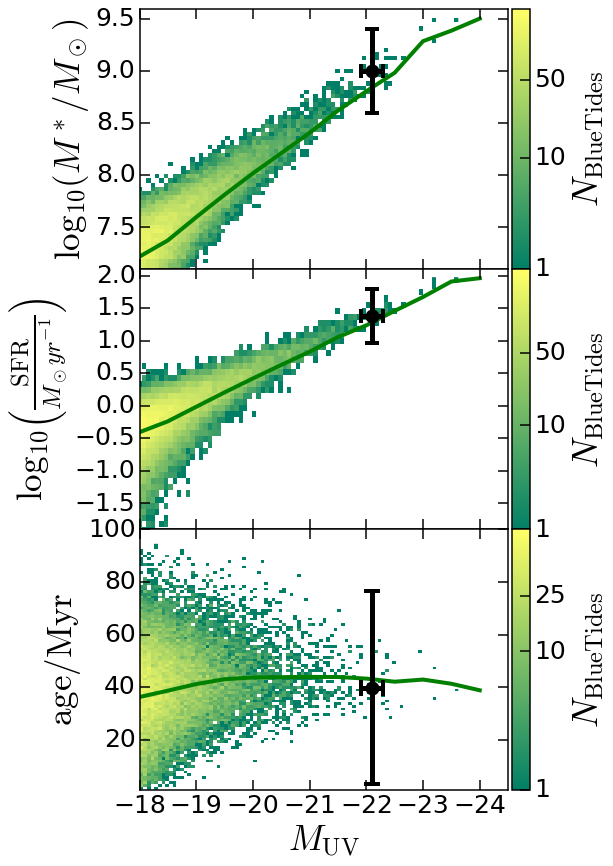}
\caption{Stellar mass, star formation rate and stellar ages versus UV
luminosity for the galaxies at $z=11$ in \bluetides. 
The large black data point denotes
the inferred constraints on GN-$z11$ from O16. The green line shows the mean value for \bluetides\ in UV magnitude bins. The 2D histogram shows the distribution of galaxies in \bluetides.}
\label{fig:properties}
\end{figure}


\section{Properties: Comparison with observations}\label{sec:properties}

\subsection{UV Continuum Slope}\label{sec:properties.SED.UVCS}

The current lack of deep mid-IR imaging limits observations of
GN-$z11$ to the measurement of the rest-frame UV continuum leaving
only the UV continuum slope $\beta$ as a spectral diagnostic. The
slope inferred from the O16 observations is $-2.5\pm 0.2$, which is
broadly consistent with the intrinsic UV continuum slope predicted by
\bluetides\ for bright galaxies at $z=11$ ($\beta_{\rm int}\approx
-2.6$), suggesting that GN-$z11$ has little or no dust
attenuation. This can be seen in Figure \ref{fig:SED} where we show
the average intrinsic SED of bright ($M_{\rm UV}\approx -22$) galaxies at $z=11.1$. We note, however, that the predicted
intrinsic UV continuum slope is also sensitive to the choice of SPS model, 
IMF, and assumed Lyman continuum escape fraction. Alternative 
choices can result in bluer intrinsic UV continuum slopes leaving open the 
possibility of some dust attenuation (see Wilkins et al. 2016).

\subsection{Stellar mass, SFR and stellar ages}\label{sec:properties.SF}

\bluetides\ makes predictions for a number of properties of GN-$z11$ 
which have been
inferred by O16. In
Figure~\ref{fig:properties} we show the stellar masses, star formation
rates and stellar ages as a function of UV luminosity for the $z=11$
galaxies in \bluetides. The black data points show the corresponding
values inferred for GN-$z11$  by O16. We can see that in the 
relevant magnitude range, \bluetides\
predicts stellar masses $\sim 10^{9} \Msun$, SFR of a few tens
$\Msun$yr$^{-1}$ and stellar ages of about $20-60$ Myr for galaxies with
$M_{\rm UV} \sim -22$. These values are fully consistent with 
the observational constraints by O16.

\begin{figure*}
\centering
\hspace{-2cm}
\hbox{
\hspace{-3cm}
\includegraphics[width=55pc]{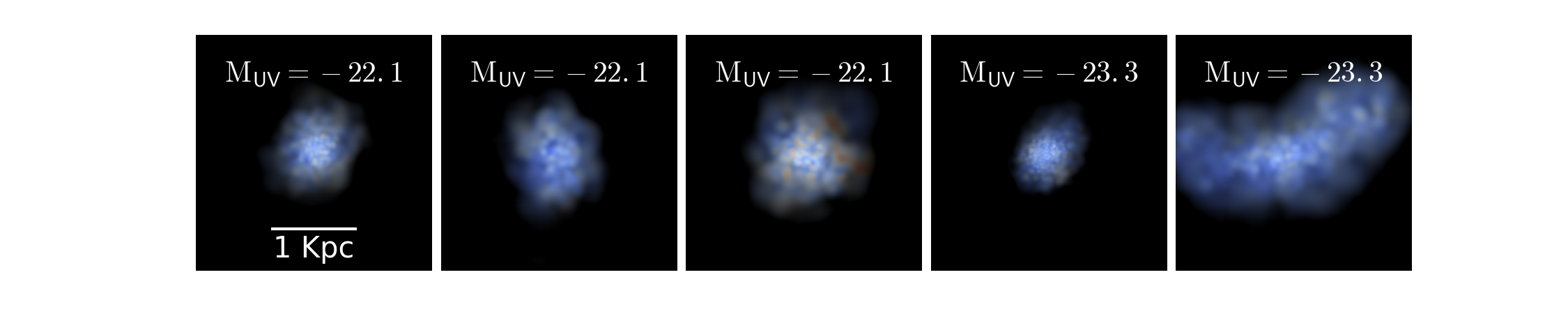}}
\caption{The stellar surface density color coded by stellar age (blue to red)
  for a sample of $M_{\rm UV} <-22$ galaxies selected from the \bluetides\ simulation
at $z=11$. We show three galaxies most closely matching the magnitude of
GN-$z11$ and the two  brightest galaxies in \bluetides\ at $z=11$.}
\label{fig:galimages}
\end{figure*}

\section{Properties: Predictions}

As galaxies with the presently observed characteristics
of GN-$z11$ exist in \bluetides\ it is useful to investigate their
other properties. \bluetides\ has high enough
spatial  resolution (180 pc at $z=11$) to allow determination
of galaxy morphologies (see Feng et al. 2015). The
simulation also tracks
gas and stellar metallicities and includes modeling of black holes. Here 
we make predictions for these aspects.




\subsection{Morphologies}\label{sec:predicted_properties.morph}

In Figure~\ref{fig:galimages} we show the stellar surface density (for
a random orientation) for a sample of five galaxies with $M_{\rm UV} <
-22$ in the $z=11$ snapshot of the \bluetides\ simulation. Three of the
galaxies closely match the brightness of GN-$z11$ (on the left) and two
are examples of brighter galaxies. Even though massive and bright, the
galaxies show irregular, disturbed morphologies and have typical sizes
of $\sim$1~kpc.  Note that in Feng et al. 2015 we found from a visual
and kinematic analysis that the most massive galaxies at $z=8$  are 
nearly all classified as disks. We can see here that this does not appear to 
be the case as early as $z=11$.

\subsection{Metallicity}\label{sec:predicted_properties.Z}

\begin{figure}
\centering
\includegraphics[width=20pc]{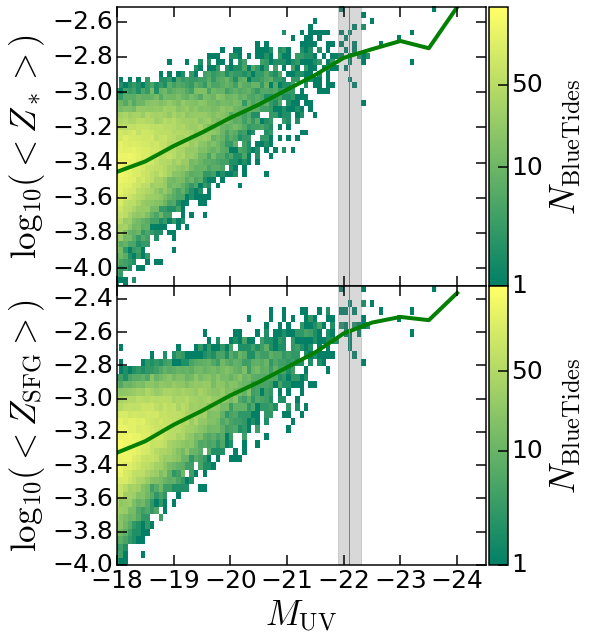}
\caption{Mean stellar and star forming gas metallicity versus UV luminosity for galaxies at $z=11$ in \bluetides. The green line shows the mean value for \bluetides\ in UV magnitude bins. The 2D histogram shows the distribution of galaxies in \bluetides. The shaded bands show the luminosity of GN-$z11$
and its observational uncertainty.}
\label{fig:properties_observed}
\end{figure}

In Figure \ref{fig:properties_observed} we show predictions for both
the star forming gas and stellar metallicity of galaxies at $z=11$ in
\bluetides. Galaxies in the simulation follow a strong luminosity -
metallicity relationship. For bright galaxies such as GB$-z11$ we predict
mean stellar metallicities of 
$(0.05-0.1)Z_{\odot}$ with the metallicity of star forming gas being 
about a factor of 2 higher.

\subsection{AGN contribution and black hole masses}\label{sec:predicted_properties.AGN}

\begin{figure}
\centering
\includegraphics[width=20pc]{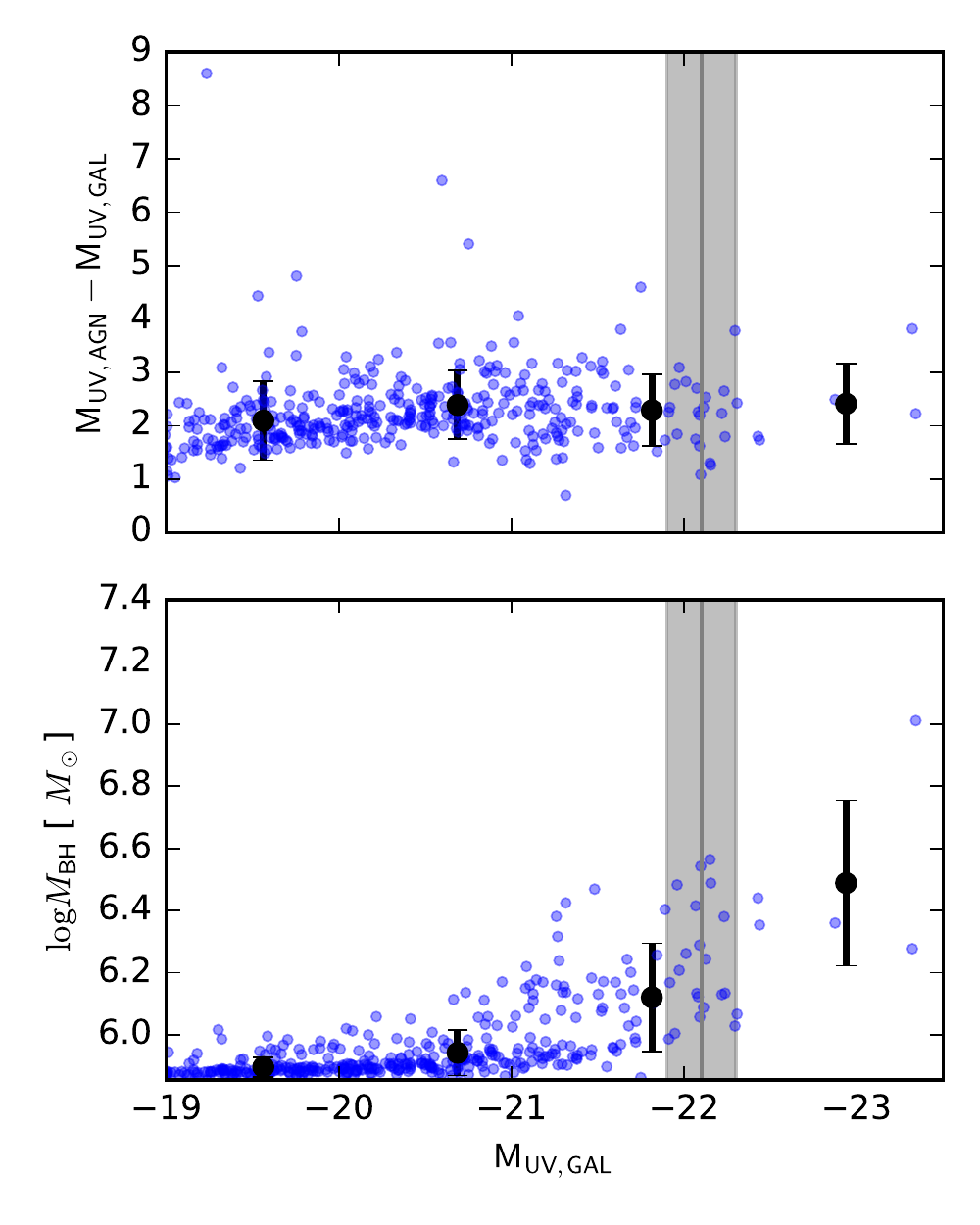}
\caption{Top panel: Prediction for the magnitude difference 
between the AGN and its host galaxy for galaxies and black holes in
\bluetides\ at $z=11$. Bottom panel: the predicted black hole masses as a
function of host galaxy magnitude. Black points denote the mean and standard deviation in UV magnitude bins. The gray area shows the region corresponding
to the magnitude of GN-$z11$.}
\label{fig:SMBH}
\end{figure}

In the top panel of Figure \ref{fig:SMBH} we show the prediction for the magnitude
difference between the active galactic nuclei (AGN) and 
their host galaxies in \bluetides\ at
$z = 11$ (see Feng et al. 2016 for details on AGN LFs; see also, Di Matteo et al. 2012; Di Matteo et al. 2016, in prep). Overall the AGN are faint (1-2 magnitudes fainter than their
hosts) at this redshift and do not contribute significantly to the
observed UV luminosity. We do not predict that an AGN should be
responsible for an enhancement in the observed GN-z11 UV flux. 

The  corresponding
black hole mass as a function of galaxy UV magnitude
 is shown in the bottom panel of Figure \ref{fig:SMBH}.
For host galaxies with $M_{\rm UV} \sim -22$, \bluetides\ predicts a 
population of massive black holes in the 
range of $M_{\rm BH} = 10^{6-7}\,{\rm  M_{\odot}}$.

\section{Conclusions}\label{sec:c}

Using the large \bluetides\ cosmological hydrodynamical simulation we
have studied the properties of the population of rare, bright galaxies at
$z=11$. As demonstrated by O16 this population is
accessible using observations from {\em Hubble} and {\em Spitzer}. Our
conclusions are summarized as follows:

\begin{itemize}

\item Within the \bluetides\ volume we find $\sim 30$ galaxies with
  $M_{\rm UV}< -22$ at $z=11$, implying a space density of $\sim
  2\times 10^{-7}/{\rm Mpc^{3}}$. Within the volume probed by O16 we expect approximately $0.17$ objects. The probability
  of observing one or more objects is then approximately $13$ per cent according to \bluetides\ and
  thus consistent with the discovery of GN-$z11$.

\item Galaxies in the simulation with luminosities similar to GN-$z11$
  have stellar masses, star formation rates, and stellar ages similar 
  to those inferred for GN-$z11$.

\item The observed SED of GN-$z11$ closely matches the intrinsic SED
  predicted by \bluetides\, suggesting that GN-$z11$ has little or no
  dust attenuation. However, the predicted SED is sensitive to the
  choice of SPS model, IMF,and Lyman continuum 
  escape leaving open the possibility of a bluer
  intrinsic slope and thus some moderate dust attenuation.

\item Bright galaxies at $z=11$ are predicted to harbor faint AGN
  accounting for $<20$ per cent of the total UV luminosity with with masses $\sim 10^{6-6.6}
  \Msun$ 

\end{itemize}

We find that 
the $\Lambda$CDM predictions that \bluetides\ represent are
consistent with this highest redshift spectroscopically
confirmed galaxy. This is significant success of the model, given
that simple extrapolations from lower redshift observations would 
otherwise  make GN-$z11$ an extreme outlier.
\bluetides\ also predicts that brighter galaxies
exist at this redshift and also 
that galaxies of similar luminosities exist at higher redshift.
These are much rarer (for example, only 5 galaxies 
with $M_{\rm UV} < -22 $ are present in the simulation volume
at $z=12$). The enormous increase in volume surveyed by
the  \wfirst\ mission (Spergel, Gehrels, Breckinridge 2013)
will however make even these galaxies accessible. If our model is
correct, the \wfirst\
High-Latitude Survey, with its sky area of 2200 deg$^{2}$ should 
detect about $1000$ galaxies (brighter than $M_{\rm UV} < -22$) from $z=11$ out to $z\sim 13.5$ (Waters et al. 2016). Such objects are likely to be seen during the early stages of
reionization, and will truly be monsters in the dark.

\subsection*{Acknowledgements}

We thank Pascal Oesch for a discussion on the recent observational result.
We acknowledge funding from NSF ACI-1036211,  NSF AST-1517593, NSF AST-1009781, and the BlueWaters PAID program. The
\bluetides\ simulation was run on facilities on BlueWaters at the
National Center for Supercomputing Applications. SMW acknowledge
support from the UK Science and Technology Facilities Council.
DN acknowledges support from NSF grant AST-1412768 and 
the Research Corporation.

\label{lastpage}
\end{document}